\documentclass[pre,twocolumn,floatfix,superscriptaddress]{revtex4}

\usepackage{amsmath}
\usepackage{amssymb}
\usepackage{graphicx}

\begin{document}

\title{Nonlinearity of Mechanochemical Motions in Motor Proteins}

\author{Yuichi Togashi}
\email{togashi@cmc.osaka-u.ac.jp}
\altaffiliation{Present affiliation: Applied Information Systems Division, Cybermedia Center, Osaka University, Ibaraki, Osaka, Japan}
\affiliation{Nanobiology Laboratories, Graduate School of Frontier Biosciences, Osaka University, Suita, Osaka, Japan}
\author{Toshio Yanagida}
\affiliation{Nanobiology Laboratories, Graduate School of Frontier Biosciences, Osaka University, Suita, Osaka, Japan}
\author{Alexander S. Mikhailov}
\affiliation{Department of Physical Chemistry, Fritz Haber Institute of the Max Planck Society, Berlin, Germany}
\date{May 7, 2010}

\maketitle

\section*{Abstract}

The assumption of linear response of protein molecules to thermal
noise or structural perturbations, such as ligand binding or
detachment, is broadly used in the studies of protein dynamics.
Conformational motions in proteins are traditionally analyzed in terms
of normal modes and experimental data on thermal fluctuations in such
macromolecules is also usually interpreted in terms of the excitation
of normal modes. We have chosen two important protein motors --- myosin
V and kinesin KIF1A --- and performed numerical investigations of their
conformational relaxation properties within the coarse-grained elastic
network approximation. We have found that the linearity assumption is
deficient for ligand-induced conformational motions and can even
be violated for characteristic thermal fluctuations.
The deficiency is particularly
pronounced in KIF1A where the normal mode description fails completely
in describing functional mechanochemical motions. These results
indicate that important assumptions of the theory of protein dynamics
may need to be reconsidered.
Neither a single normal mode, nor a superposition of such modes
yield an approximation of strongly nonlinear dynamics.


\section*{Author Summary}

Biological cells use a variety of molecular machines representing enzymes,
ion channels or pumps, and motors.
Motor proteins are nanometer-size devices generating forces and actively
moving or rotating under the supply of chemical energy through ATP
hydrolysis.
They are crucial for many cell functions and promising for
nanotechnology of the future.
Although such motors represent single molecules, their operation cycles
cannot be in detail followed in simulations even on the best modern
supercomputers and some approximations need to be employed.
It is often assumed that conformational dynamics of motor proteins
is well described within a linear response approximation and corresponds
to excitation of normal modes.
We have checked this assumption for two motor proteins, myosin V and
kinesin KIF1A.
Our results show that, while both these biomolecules respond by
well-defined motions to energetic excitations, these motions are
essentially nonlinear.
The effect is particularly pronounced in KIF1A where relaxation proceeds
through a sequence of qualitatively different conformational changes,
which may facilitate complex functional motions without additional
control mechanisms.

\section*{Introduction}

\begin{figure*}
\begin{center}
\includegraphics[width=173.48mm]{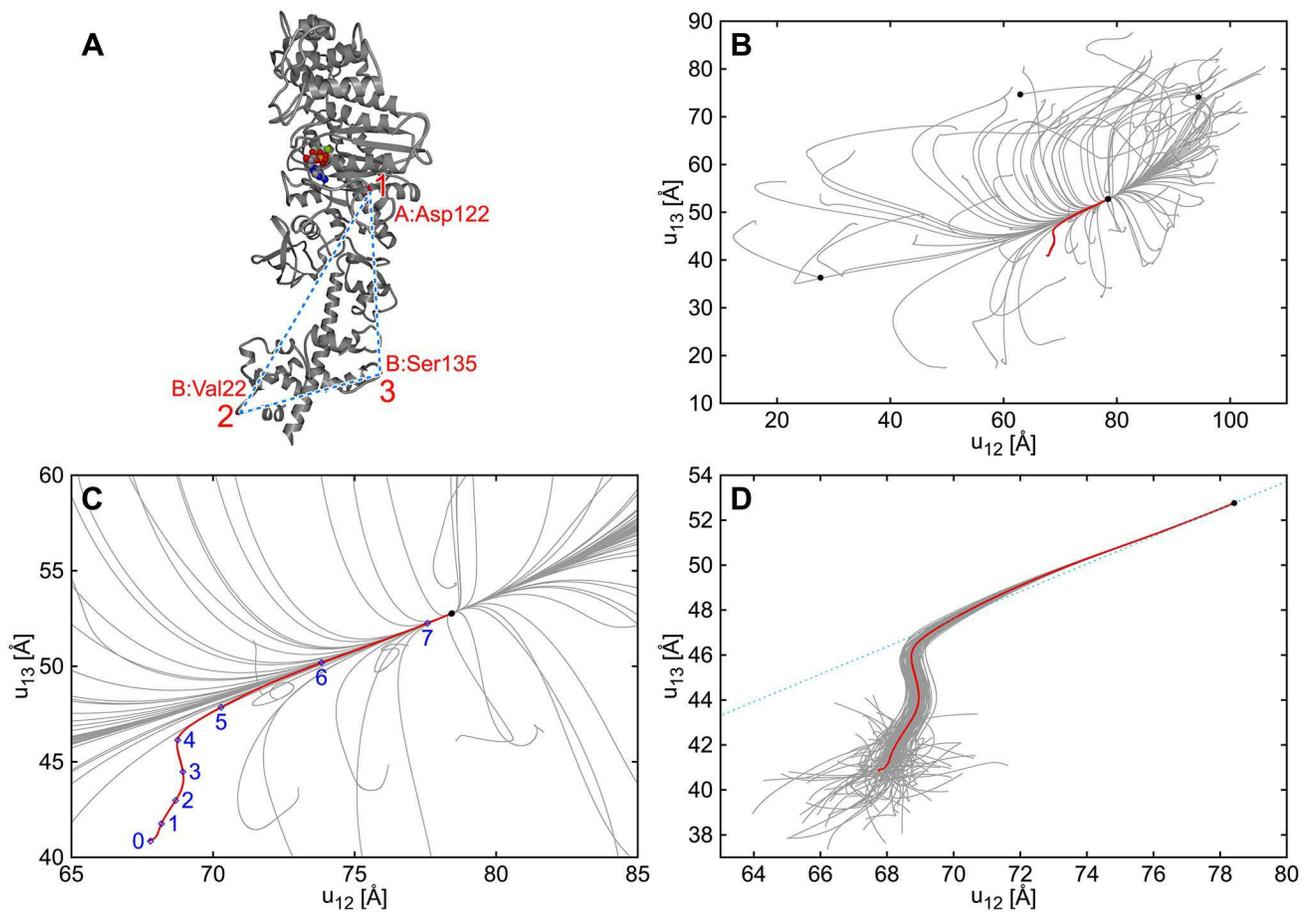}
\end{center}
\caption{
{\em Relaxation paths of myosin V.}
The elastic network model is constructed for the ATP-bound structure
as the reference state.
The red line shows the trajectory in the plane of distances
$u_{12}$ and $u_{13}$ between labels
$1$, $2$ and $3$ indicated in panel (A) starting from
the nucleotide free state, so that this path
corresponds to the conformational transition upon ATP binding.
In panels (B) and (C) (magnified), gray lines display trajectories
starting from 100 different random initial conditions (see Methods).
In panel (D), gray lines represent relaxation trajectories with
100 different random deformations applied to the same initial structure
as that for the red line.
The dotted line in panel (D) shows the direction of distance changes
corresponding to the slowest normal relaxation mode.
Black dots indicate (meta)stable states reached.
Times $t_{i}$ at points $i=0\ \mathrm{to}\ 7$ indicated in panel (C) are
$t_{i} = 0,\ 10,\ 30,\ 100,\ 300,\ 1000,\ 3000,\ \mathrm{and}\ 10000$, respectively.
}
\label{Fig-Myo5}
\end{figure*}

\begin{figure*}
\begin{center}
\includegraphics[width=134.11mm]{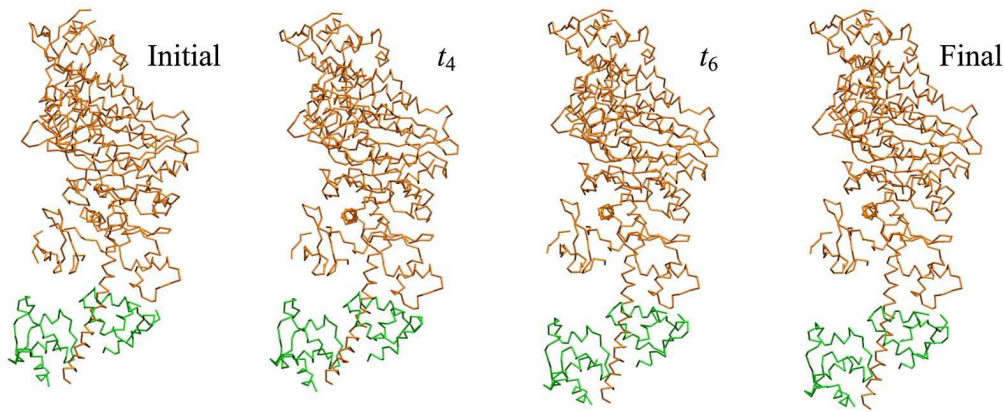}
\end{center}
\caption{
{\em Relaxation motion in myosin V.}
Snapshots of the conformations of myosin V along the special relaxation path
are shown.
The essential light chain, displayed in green, is included into the model.
}
\label{Fig-snapshots-Myo5}
\end{figure*}

Protein machines, which may represent enzymes, ion pumps or molecular motors,
play a fundamental role in biological cells and understanding of their
activity is a major challenge.
Operation of these machines is based on slow conformational motions
powered by external energy supply, often with ligands (such as ATP).
In molecular motors, binding of ATP and its subsequent hydrolysis induce
functional mechanochemical motions, essential for their operation.
These motions, which follow after an energetic activation,
are conformational relaxation processes.

Large-scale conformational changes may take place in proteins
as a result of ligand binding \cite{MolMovDB}.
Despite the large magnitude of such changes,
they are nonetheless often considered in the framework of
the linear response theory \cite{KideraLRT} and
the normal mode approximation \cite{GoNMA1983,Brooks1983,Baharbook,Karplus,Levitt}.
The normal mode analysis is furthermore broadly employed in
the elastic-network studies of
proteins \cite{Tirion,Hinsen1998,Doruker,Atilgan2001,Bahar1997,Haliloglu,Tama,Liao,Bahar,Baharbook,LeiYang,Zheng}.
However, there is no general justification to assume that
relaxation processes in proteins are linear and
this assumption has to be verified for particular macromolecules.

It is known that relaxation processes in complex dynamical systems
may be strongly nonlinear and deviate much from simple exponential
relaxation.
As an example borrowed from a distant field, we can mention
the Belousov-Zhabotinsky reaction which exhibits a great variety of
spatiotemporal patterns (pacemakers, rotating spiral waves)
that are however only complicated transients
accompanying relaxation to the equilibrium state \cite{BZWave,BZSpiral}.
Conformational relaxation in single protein molecules may also
be a complicated process, comprising qualitatively different
kinds of mechanochemical motions.

While partial unfolding and refolding, associated with ligand binding,
are known for some protein machines,
such as the enzyme adenylate kinase \cite{Miyashita},
usually functional conformational motions
in molecular machines and, specifically, in motor proteins are {\em elastic}.
This means that the pattern of contacts between the residues in a protein
is not changed upon ligand binding and preserved during
the relaxation process, as generally assumed in the elastic network
modeling (ENM).

Here, we provide detailed analysis of conformational relaxation processes,
associated with ligand binding and hydrolysis, in two motor proteins
--- myosin V \cite{MyosinBook,MyosinClass} and kinesin KIF1A \cite{KIFreview}.
Our investigations, performed in the framework of
the ENM approximation, reveal that nonlinearity is characteristic
for both macromolecules and the normal mode description is not really
applicable for any of them.
For KIF1A, a monomeric motor protein from the kinesin superfamily,
nonlinear effects are found to dominate completely functional
mechanochemical motions which turn out to be qualitatively different
from the normal mode predictions.
Despite the nonlinearity, well-defined conformational relaxation paths,
robust against perturbations, have been found in both motor proteins.


\begin{figure*}
\begin{center}
\includegraphics[width=173.48mm]{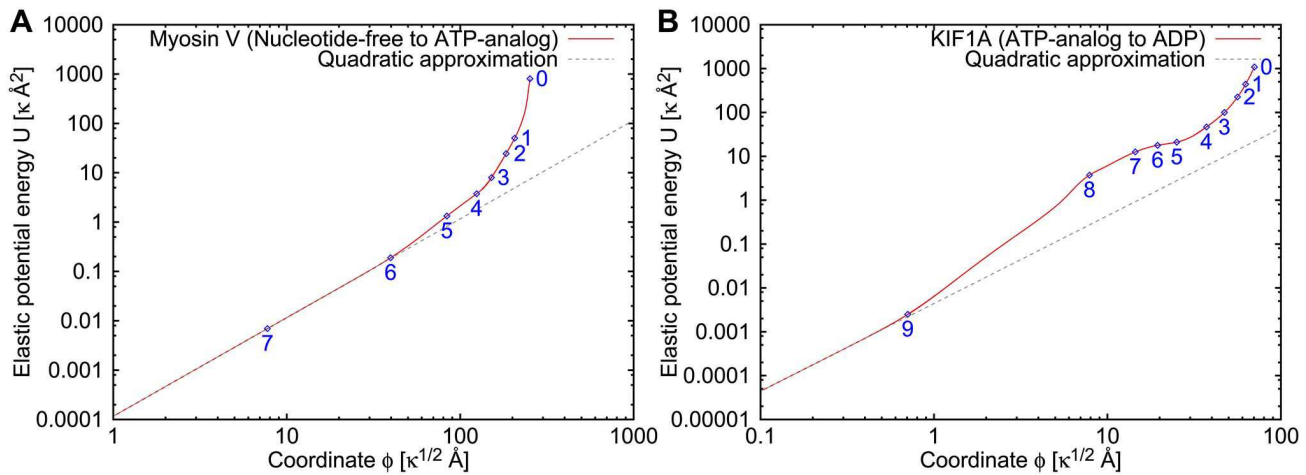}
\end{center}
\caption{
{\em Profiles of elastic potential energy.}
(A) Myosin V and (B) KIF1A.
Elastic energy $U$ during transitions,
corresponding to the trajectories shown by red lines in
Figures \ref{Fig-Myo5} and \ref{Fig-KIF1A},
is plotted against coordinate $\phi$.
The dashed line shows $U = \frac{1}{2} \lambda_{1} \phi^{2}$,
the quadratic approximation corresponding to
the slowest normal mode (see Methods).
Numbers correspond to time moments indicated in
Figures \ref{Fig-Myo5}C and \ref{Fig-KIF1A}C.
}
\label{Fig-pot}
\end{figure*}

\begin{figure*}
\begin{center}
\includegraphics[width=173.48mm]{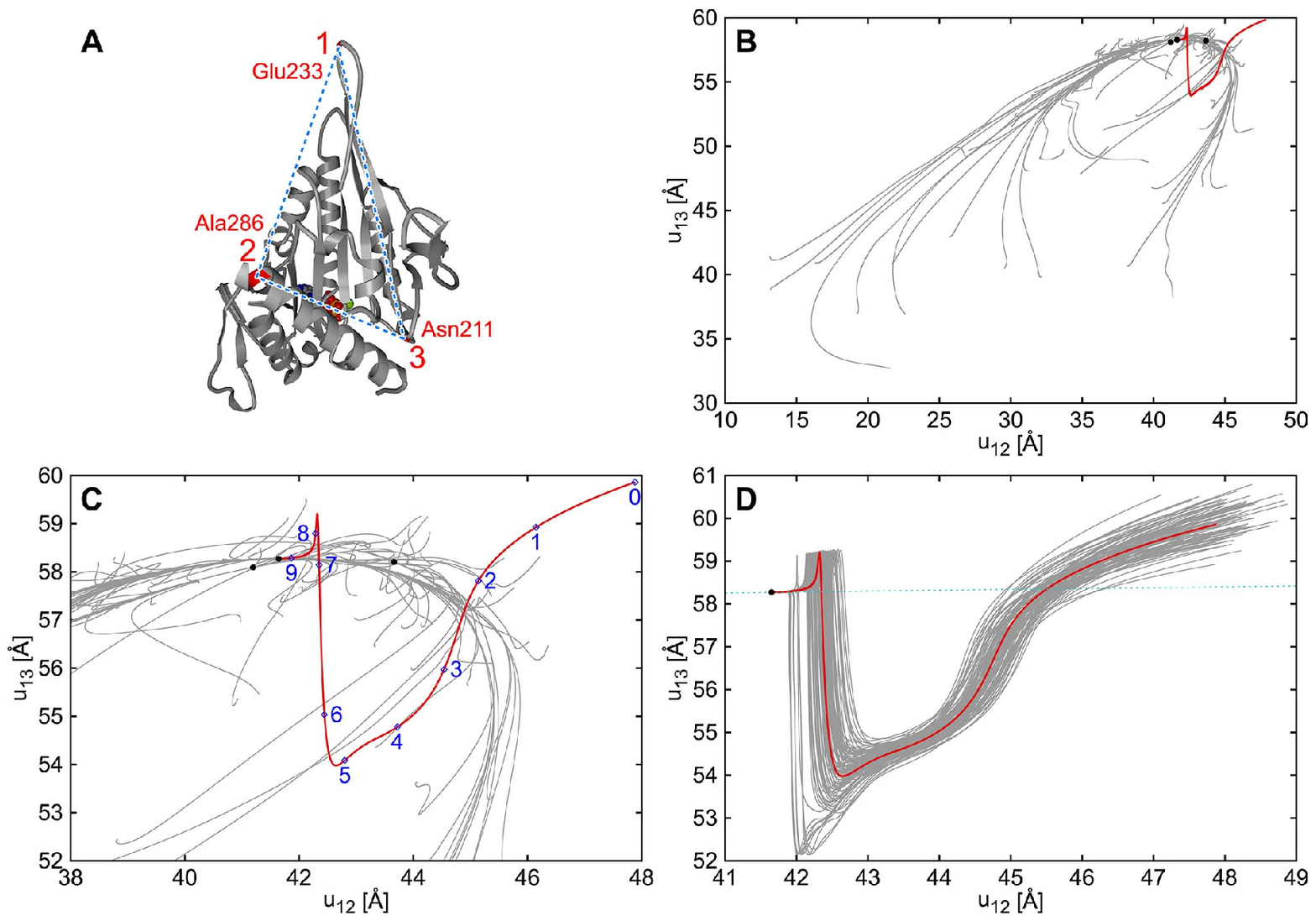}
\end{center}
\caption{
{\em Relaxation paths of KIF1A.}
The ADP-bound structure has been used to construct the elastic network.
The visualization labels are indicated in panel (A), and
the relaxation paths are displayed in panels (B) to (D) in the same way
as in Figure \ref{Fig-Myo5}, panels (B) to (D), respectively.
The initial condition for the red line is the ATP-bound state,
so that this trajectory corresponds to the transition from
the ATP-bound state to the ADP-bound state.
Times $t_{i}$ at points $i=0\ \mathrm{to}\ 9$ indicated in panel (C) are
$t_{i} = 0,\ 0.1,\ 0.3,\ 1,\ 3,\ 10,\ 20,\ 25,\ 30,\ \mathrm{and}\ 100$, respectively.
}
\label{Fig-KIF1A}
\end{figure*}

\section*{Results}

Within the coarse-grained ENM approach, a protein is modeled as
a network of point-like particles, corresponding to residues,
which are connected by a set of elastic links \cite{Hinsen1998,Doruker}.
A link between two particles is present if the distance between them
in the equilibrium conformation of the considered macromolecule is
shorter than a cutoff length.
The elastic energy of the network is
$U = (\kappa/2) \sum_{ij} A_{ij} \left( d_{ij} - d_{ij}^{(0)} \right)^{2}$,
where $\kappa$ is the stiffness constant of the network links,
$A_{ij}$ is the matrix of connections inside the network,
$d_{ij}$ is the distance between particles $i$ and $j$, and
$d_{ij}^{(0)}$ is the respective distance in the equilibrium reference state.
The characteristic time scales of functional mechanochemical motions
in motor proteins are in the millisecond range and slow conformational
relaxation motions on such timescale should be overdamped \cite{Kitao1991}.
Neglecting hydrodynamic interactions,
relaxation dynamics is then described by equations
$d \mathbf{R}_{i} / dt = - \Gamma \partial U / \partial\mathbf{R}_{i}$ for
the coordinates $\mathbf{R}_{i}$ of the particles,
where $\Gamma$ is their mobility.
Relaxation dynamics for elastic networks of proteins
has been previously considered \cite{Piazza2005}.

Despite a wide-spread misunderstanding,
elastic dynamics is generally nonlinear.
For example, macroscopic objects, such as ribbons or membranes,
can still exhibit pronounced nonlinear effects of spontaneous
twisting or buckling, while fully retaining their elastic behavior
and not undergoing plastic deformations \cite{LandauElasticity}.
The energy $U$ of an elastic network is quadratic in terms of
the distances $d_{ij}$ and the forces acting on the particles are
linear in terms of such distances.
However, the distance
$d_{ij} = \left\vert \mathbf{R}_{i} - \mathbf{R}_{j} \right\vert$ is
itself a {\em nonlinear function} of the coordinates
$\mathbf{R}_{i}$ and $\mathbf{R}_{j}$ and this makes
the forces also nonlinear functions of dynamical variables.
The presence of nonlinear effects in conformational
relaxation of proteins in the ENM approximation has been previously
demonstrated \cite{Hayashi2007,Togashi2007}.

Explicitly, relaxation dynamics of considered proteins is
described by equations (\ref{relaxation}) in the Methods section,
where further details are also given.
To study conformational relaxation, these equations were numerically
integrated starting from various initial conditions.

\subsection*{Myosin V}

The reference conformation, used to construct the elastic network, was
that of the ATP(analog) bound state
(Protein Data Bank (PDB) ID code: 1W7J,
with MgADP-BeFx as the ATP analog \cite{1W7J}).
As the initial condition, the conformation corresponding to the
nucleotide-free state was taken (PDB ID: 1OE9 \cite{1OE9}).
The elastic network had 855 particles connected by 7261 links.
Note that only the residues whose $\alpha$-carbon positions are
contained in both PDB data sets have been taken to construct
the network.
Additionally, relaxation processes starting from randomly generated
initial conditions (see Methods) have been considered.
For visualization purposes, motions of three particles
(Asp122 in chain A, and Val22 and Ser135 in chain B) have been
traced (Figure \ref{Fig-Myo5}A).
Thus, each relaxation process was characterized by a certain trajectory
in the space of distances between the three chosen labels.

Figures \ref{Fig-Myo5}B,\ref{Fig-Myo5}C display 100 conformational relaxation
trajectories, each starting from a different random initial condition.
Although the initial conditions were generated by applying relatively
strong deformations (without unfolding) to the reference state,
almost all of them were leading back to that reference state,
with just a few metastable states found.
Furthermore, one can observe that the trajectories converge to
a well-defined relaxation path.

The red trajectory in Figures \ref{Fig-Myo5}B,\ref{Fig-Myo5}C is for the relaxation
starting from the nucleotide-free conformational state of myosin V
(so that the mechanochemical motion following ATP binding is simulated).
After a transient, this special trajectory joins the well-defined
relaxation path.
This functional trajectory is robust against perturbations,
as shown by Figure \ref{Fig-Myo5}D.
Several snapshots of the conformation along this trajectory are shown
in Figure \ref{Fig-snapshots-Myo5} (see also Video S1).

The attractive path corresponds to a deep energy valley in
the energy landscape of myosin V.
Once this valley is entered, the conformational relaxation motion
becomes effectively one-dimensional and characterized by
a single mechanical coordinate.
The profile of the elastic energy along the bottom of such energy valley
determines the dependence of the elastic energy on the collective
mechanical coordinate (see Methods).

Figure \ref{Fig-pot}A shows the dependence of the elastic energy
along the special attractive relaxation path starting from the nucleotide-free state
and leading to the ATP-bound state.
Markers indicate positions along the trajectory
in Figure \ref{Fig-Myo5}C.
For $t > t_{6}$, the elastic energy $U$ is approximately quadratic
in terms of the mechanical coordinate $\phi$,
i.e. $U(\phi) \simeq (\lambda /2) \phi^{2}$.
Because $d\phi / dt = - dU / d\phi$, this implies that then
$d\phi / dt = - \lambda\phi$ and the relaxation is exponential.
Only within such harmonical neighborhood of the reference state,
the normal mode description becomes applicable
(see Methods for further discussion).

The dotted blue line in Figure \ref{Fig-Myo5}D shows the direction of
the distance changes corresponding to the slowest normal mode (see Methods).
The nucleotide-free state of myosin V lies
away from this direction and also outside of the harmonical neighborhood.
The initial stage of the functional mechanochemical motion
(until time $t_{6}$) cannot be quantitatively analyzed
in terms of the normal modes.

\begin{figure*}
\begin{center}
\includegraphics[width=173.23mm]{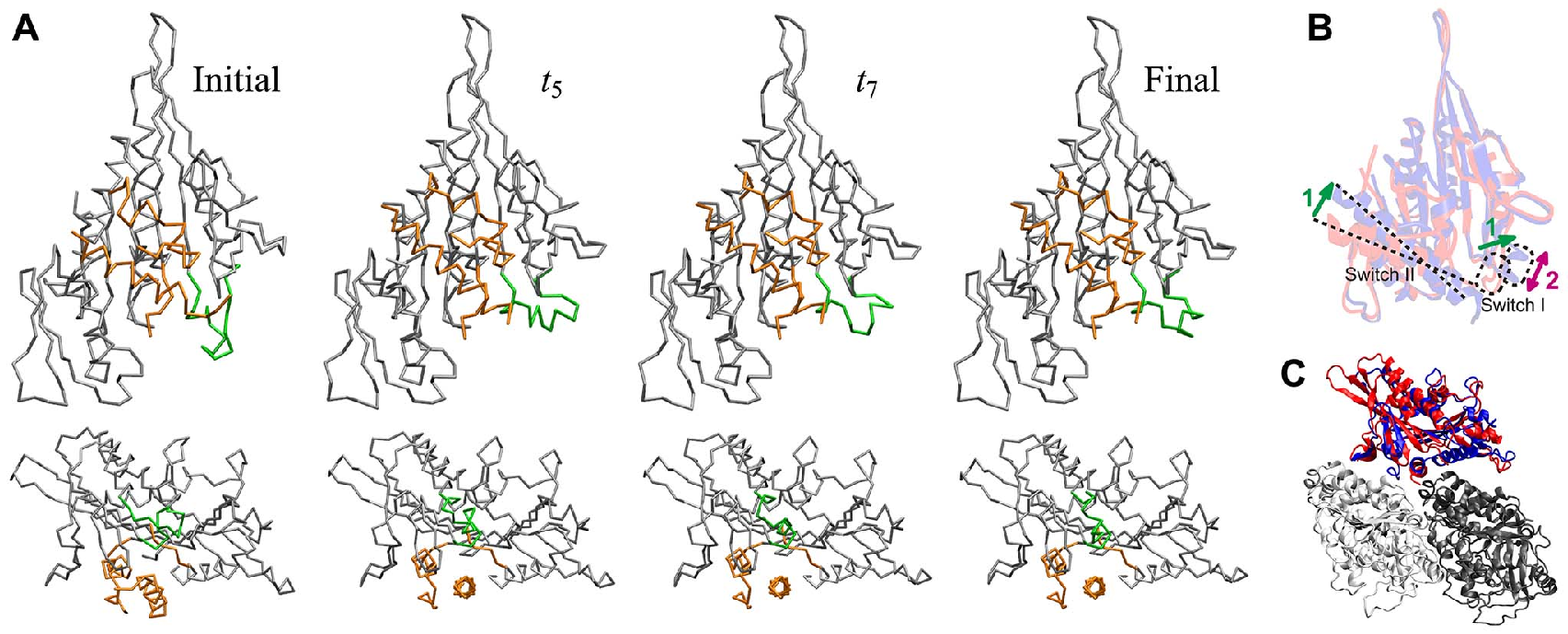}
\end{center}
\caption{
{\em Relaxation motion in KIF1A.}
(A) Conformation snapshots (seen from two different viewpoints).
Switch I and switch II regions are indicated in green and
orange, respectively.
(B,C) Schematic representation of the relaxation motion,
observed in the simulation from the ATP-bound state (red)
to the ADP-bound state (blue).
In switch I region, as shown in panel (B), reconfiguration of
the structure (2), i.e., transformation from a loop to
an $\alpha$-helix, occurs only after the sliding motion (1) is
completed.
For reference, relative positions of KIF1A and tubulin monomers
(PDB ID: 2HXF and 2HXH \cite{2HXF-2HXH}) are shown in panel (C).
}
\label{Fig-snapshots-KIF1A}
\end{figure*}

\begin{figure}
\begin{center}
\includegraphics[width=82.97mm]{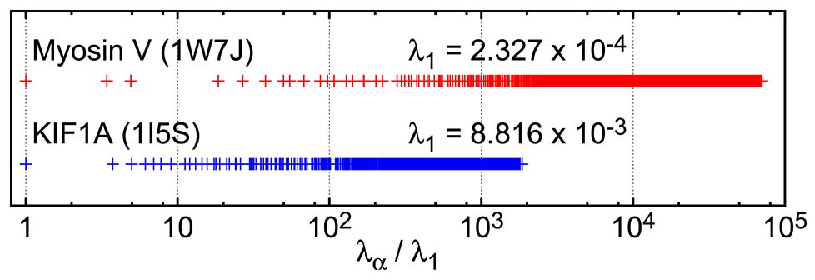}
\end{center}
\caption{
{\em Eigenvalue spectra of the elastic network models.}
Eigenvalues $\lambda_{\alpha}$ in the normal mode description
(eqn. (\ref{eigen})),
normalized to the lowest non-zero eigenvalue $\lambda_{1}$,
are shown.
There is a significant gap between the lowest and
the second lowest modes.
}
\label{Fig-eigen}
\end{figure}

\subsection*{Kinesin KIF1A}

The reference conformation for KIF1A is the ADP-bound state
(PDB ID: 1I5S, with MgADP \cite{1I6I-1I5S}).
Relaxation starting from the initial
condition, corresponding to the ATP(analog)-state
(PDB ID: 1I6I, with MgAMPPCP as an ATP analog \cite{1I6I-1I5S})
and from randomly generated initial conditions was considered.
The elastic network has 320 particles and 2871 links.
Only the residues whose $\alpha$-carbons are in both
PDB data sets have been used.
Visualization labels are Glu233, Ala286, and Asn211
(Figure \ref{Fig-KIF1A}A).

100 relaxation trajectories, starting from random initial conditions,
are shown in Figures \ref{Fig-KIF1A}B,\ref{Fig-KIF1A}C.
The presence of an attractive relaxation path,
corresponding to a deep energy valley, can be noticed.

The red lines in Figures \ref{Fig-KIF1A}B,\ref{Fig-KIF1A}C display the special
relaxation trajectory starting from the ATP-bound state.
Surprisingly, we find that, in contrast to myosin V,
this trajectory is {\em different} from the typical relaxation path.
By applying small random initial perturbations to the initial
ATP-bound state and integrating the dynamical equations,
it can be demonstrated that this trajectory is, however,
also stable with respect to the perturbations (Figure \ref{Fig-KIF1A}D).
The dotted blue line in Figure \ref{Fig-KIF1A}D shows the direction of
the distance changes in the slowest normal mode of KIF1A.

Thus, in KIF1A the deep energy valley leading to the reference
ADP-bound state gets branched at some distance from it.
The path corresponding to the functional mechanochemical motion
from the ATP-bound state belongs to the side branch.
Only at the final relaxation stage,
in the immediate vicinity of the equilibrium,
the valleys merge and the functional motion begins to coincide
with the typical relaxation motion in this protein.

The branching of the energy valley is already an indication of
strong nonlinearity in the relaxation dynamics.
We have also determined the profile of the elastic energy $U$ as
a function of the mechanical coordinate $\phi$ along the path
connecting the ATP- and ADP-bound states (Figure \ref{Fig-pot}B).
The profile becomes quadratic only starting from time $t_{9}$,
very close to the equilibrium reference state.

Figure \ref{Fig-snapshots-KIF1A} shows snapshots of KIF1A
along the special attractive relaxation path (see also Video S2).
At the early relaxation stage (until $t_{5}$),
the relaxation motion represents a combination of the rotation of
the switch II helix and of the sliding of the switch I loop.
Relaxation at the end of such initial stage is
apparently hindered, as revealed in the presence of a plateau
in the dependence of the elastic energy on the mechanical coordinate
in Figure \ref{Fig-pot}B near $t = t_{5}$.
Only once the sliding is completed,
further local structural reorganization,
representing a transition from the loop to the $\alpha$-helix,
becomes possible and is indeed observed approximately after time $t_{6}$.

\section*{Discussion}

The normal mode description is broadly used in structural studies of proteins.
The analysis of thermal fluctuations and the interpretation of
the respective experimental structural data are traditionally performed
assuming that fluctuations are
linear and, hence, correspond to thermal excitation of various normal modes
(see, e.g., \cite{GoNMA1983,Brooks1983}).
The linear response of a protein macromolecule to structural perturbations,
such as ligand binding, is an often used assumption \cite{KideraLRT}.
To a large extent, the elastic-network analysis of
ligand-induced macromolecular motions is based on determining normal modes in
the elastic networks of the considered proteins (see, e.g., \cite{Baharbook}).
The patterns of atomic displacements in such normal modes are further compared
with the experimentally measured atomic displacements in the same proteins
that are induced by a change of the chemical state,
such as binding of an ATP molecule \cite{Tama,Liao,Bahar,Baharbook,LeiYang}.
Large overlaps with only
a few slowest normal modes are seen as the evidence for the applicability of
the elastic-network ansatz,
whereas the wide distribution of overlaps is considered as the indication
that the elastic network description fails for a particular macromolecule.
Specifically, strong overlaps between ligand-induced conformational changes and
atomic displacements in the few slowest modes have been found for
scallop myosin and $\mathrm{F}_{1}$-ATPase, while such overlaps were absent for
kinesin KIF1A \cite{Zheng}.

Our numerical investigations of elastic conformational motions in two
motor proteins (myosin V and KIF1A) have revealed however that in both of
them the nonlinearities play an essential role.
While slow conformational relaxation motions in myosin
can still be qualitatively characterized
in terms of the normal modes, the normal mode description breaks down
{\em completely} for KIF1A.
The observed breakdown of the normal mode description
does not however mean that conformational motions become irregular.
We have seen that ordered and robust
mechanochemical motions are characteristic for both protein motors,
even though they cannot be described in terms of the linear response.

We want to emphasize that, when the dynamics is nonlinear,
neither a single normal mode, nor a combination of many such modes
can reproduce the motions.
Thus, the normal mode description fails completely in this case
and the problem is not that many normal modes must be taken into account.
Actually, as we have shown, even for KIF1A, one normal mode would be
sufficient to describe long-time relaxation within the harmonic domain
--- however, this domain is restricted to a tiny neighbourhood
of the equilibrium state.

Thermal fluctuations have not been explicitly included into our
dynamical ENM simulations.
However, such fluctuations are effectively generating
random conformational perturbations.
In our study, relaxation processes starting from random conformational
perturbations have indeed been considered.

In myosin V, one well-defined nonlinear conformational relaxation
trajectory, leading to the equilibrium state, has been identified.
Starting from an arbitrary initial conformation
(but still without unfolding), rapid convergence to this special
trajectory takes place.
While the motion corresponding to the special attractive trajectory
is initially nonlinear, it becomes harmonical later and
a substantial part of the ordered conformational relaxation process
is within the harmonic domain of the equilibrium state.
Similar behavior has been previously noted by us \cite{Togashi2007} for
scallop myosin and $\mathrm{F}_{1}$-ATPase,
but its detailed analysis has not yet been performed.

The situation is more complex for the monomeric kinesin KIF1A.
Instead of a unique deep energy valley leading to the
reference ADP-bound state, two such valleys, both leading to 
the equilibrium state, are present.
These valleys correspond to two kinds of ordered
conformational motions possible in the protein.

The first of them is relatively wide and, when thermal conformational
fluctuations are excited, they would typically proceed along it.
However, the conformational relaxation motion
starting from the ATP-bound state follows a different path,
which corresponds to the second energy valley branching from the
typical fluctuation path already at very small deviations
from the equilibrium state.
Note that the branching takes place as the change in the distance
between the molecular labels Glu233 and Ala286 is still less than
an angstrom, which is much smaller than the intensity of typical
thermal fluctuations for such a distance.
Thus, the nonlinear effects in KIF1A are strong even for the
typical thermal fluctuations.

Remarkably, such second relaxation path is also stable
with respect to perturbations, i.e. structurally robust.
Our numerical investigations reveal that
motion along this path can be divided into two
{\em qualitatively different} stages.
At the first of them, sliding of the switch I loop is observed,
whereas at the second stage a transition from the loop to
the $\alpha$-helix is realized.
Structural reorganization, corresponding to this transition,
is not possible until the sliding motion is completed,
lifting a restriction through the backbone chain.
Recent crystallographic studies suggest that the switch I loop/helix
plays an important role in control of the motor function
through interaction with $\mathrm{Mg}^{2+}$ and switch II \cite{Nitta2008}.

Thus, in contrast to myosin, a single ATP binding event induces in KIF1A a
complex, but ordered conformational motion characterized by two qualitatively
different consequent phases.
As we conjecture, this special dynamical property of
KIF1A may be needed for the processive motion of
this single-headed molecular motor \cite{KIFreview}.

In myosin V, conformational motions driven by random thermal
fluctuations are similar in their properties to the
relaxation motion from the nucleotide-free state.
This may facilitate exploitation of such fluctuation motions
for the motor operation, as suggested by recent
single-molecule experiments \cite{StrainSensor}.
In KIF1A, where the energy valley splits into two branches,
typical thermal conformational fluctuations are qualitatively different from
the relaxation motion starting from the previous ATP-bound state.
The latter motion is entropically hindered for thermal fluctuations
and cannot be reversed through them.
This may turn out to be important for the understanding of
the operation of the monomeric kinesin as a molecular motor.
Latest experimental techniques permit simultaneous observation of
stepping motion and conformational changes of a motor \cite{Tomishige2006}.
The coarse-grained modeling, including our present study,
can contribute further suggestions for the design
(e.g., by determining positions for fluorescent labeling)
of such experiments.

Finally, we note that our study has been based on the elastic network
approximation for proteins.
More detailed descriptions, such as, e.g., G\={o}-like models,
can also be used to consider conformational relaxation processes \cite{SwitchingGo}.
We expect that similar behavior will then be observed.

\section*{Methods}

\subsection*{Elastic network models}

In this study, we employ elastic network models where material points
are connected by a set of elastic springs \cite{Tirion,Hinsen1998,Doruker,Atilgan2001}.
Each particle corresponds to a residue in the considered protein.
The equilibrium positions $\mathbf{R}_{i}^{(0)}$ of the particles
are determined by the locations of $\alpha$-carbon atoms in
the reference state of the protein, taken from the PDB database.
Two particles in a network are connected by an elastic spring
if at equilibrium the distance
$d_{ij}^{(0)} = \left\vert \mathbf{R}_{i}^{(0)} - \mathbf{R}_{j}^{(0)} \right\vert$ between
them is less than a certain cutoff length $l_{0}$.
The natural length of an elastic link is equal to the equilibrium
distance $d_{ij}^{(0)}$.
The cutoff distance $l_{0} = 10\ \mathrm{\AA}$ has been used in our study.

The elastic forces obey the Hooke law and all springs have the same
stiffness constant $\kappa$.
Elastic torsion effects are not included.
Thus, the force acting on particle $i$ is
\begin{equation}
\mathbf{F}_{i} = -\kappa \sum_{j=1}^{N} A_{ij} \left( d_{ij} - d_{ij}^{(0)} \right) \frac{\mathbf{R}_{i}-\mathbf{R}_{j}}{d_{ij}},
\label{elasticforce}
\end{equation}
where $N$ is the total number of particles in the network,
$\mathbf{R}_{i} (t)$ is the actual position of the particle $i$ and
$d_{ij} = \left\vert \mathbf{R}_{i} - \mathbf{R}_{j} \right\vert$ is
the actual distance between two particles $i$ and $j$.
The adjacency matrix of the network is defined as having
$A_{ij}=1$ if $d_{ij}^{(0)} < l_{0}$ and $A_{ij}=0$ otherwise.
The total elastic energy of the network is
\begin{equation}
U = \frac{\kappa}{2} \sum_{i<j} A_{ij} \left( d_{ij} - d_{ij}^{(0)} \right)^{2}.
\label{potential}
\end{equation}

Because slow conformational dynamics of proteins in the solvent is
considered, the motions are overdamped (see \cite{Kitao1991})
and the velocity of a particle is proportional to the force acting
on it, i.e. $d\mathbf{R}_{i} / dt = \Gamma \mathbf{F}_{i}$ where
$\Gamma$ is the mobility.
We assume that the mobilities of all particles are the same.
Hydrodynamical effects are neglected (they can be however incorporated
into the elastic network models as shown in ref. \cite{Hydrodyn}).

Explicitly, the relaxation dynamics is described by a set of
differential equations:
\begin{equation}
\frac{d\mathbf{R}_{i}}{dt} = - \sum_{j=1}^{N} A_{ij} \left( \left\vert \mathbf{R}_{i} - \mathbf{R}_{j} \right\vert - \left\vert \mathbf{R}_{i}^{(0)} - \mathbf{R}_{j}^{(0)} \right\vert \right) \frac{\mathbf{R}_{i} - \mathbf{R}_{j}}{\left\vert \mathbf{R}_{i} - \mathbf{R}_{j} \right\vert}.
\label{relaxation}
\end{equation}
Here, time is rescaled and measured in units of $(\Gamma \kappa)^{-1}$.
Hence, the relaxation dynamics of a network is completely determined
by its pattern of connections (matrix $A_{ij}$) and the equilibrium
distances $d_{ij}^{(0)}$ between the particles.
Equations (\ref{relaxation}) were integrated
to determine conformational relaxation motions.

\subsection*{Simulations}

To prepare random initial conditions, the following procedure has been
employed.
Random static forces $\mathbf{f}_{i}$, acting on all particles in the
network have been independently generated with the constraint that
$(1/N) \sum | \mathbf{f}_{i} |^{2} = F_{ini}^{2}$.
The equations of motion were integrated in the
presence of such static forces for a fixed time $\tau_{ini}$.
The conformation which was thus reached has been then used as the initial
condition for the relaxation simulation.
The parameters were
$F_{ini} / \kappa = 0.3\ \mathrm{\AA}$, $\tau_{ini} = 10^{5} (\Gamma \kappa)^{-1}$ for myosin V and
$F_{ini} / \kappa = 0.5\ \mathrm{\AA}$, $\tau_{ini} = 3 \times 10^{4} (\Gamma \kappa)^{-1}$ for KIF1A.
With these parameter values,
relatively large overall deformations ($\sim 20\mathrm{\%}$ typical) could be reached,
while still avoiding unfolding.
In the deformed states, the lengths of the links did not exceed
$13.4\ \mathrm{\AA}$ for myosin V and $11.5\ \mathrm{\AA}$ for KIF1A.

When relaxation from specific conformations has been considered,
initial positions of all particles were allocated according to the
respective PDB structures. When robustness of a relaxation path
starting from a specific conformation was investigated, the initial
condition was prepared by randomly shifting the positions of all
particles with respect to their locations in that conformation with a
certain root-mean-square displacement $d_{ini}$.
We have chosen $d_{ini} = 2\ \mathrm{\AA}$ for myosin V and $0.5\ \mathrm{\AA}$ for KIF1A.

To visualize conformational motions, three particles
labeled as $1$, $2$ and $3$ were chosen and the distances
$u_{12}$ and $u_{13}$ were monitored in the simulations.
Thus, the relaxation motion was represented by a
trajectory on the plane $(u_{12},\ u_{13})$.

The choice of the visualization labels is essentially arbitrary.
In a simulation, motions of all residues were traced
(see, e.g., Videos S1 and S2)
and different residues could be selected for a specific visualization.
If a molecule has a low-dimensional attractive relaxation manifold,
this is a property of the respective dynamical system
and it cannot depend on the visualization method.
When selecting the labels, one should only pay attention to the fact
that the distances between them should significantly vary
during the relaxation process.
If, by chance, two labels belonging to the same stiff domain
in a protein have been taken, the distance between them would remain
almost constant, so that such a choice would be inconvenient.
When the normal mode description approximately holds and, furthermore,
relaxation is well described by a few slowest modes,
one can choose the labels so that the distances between them
reveal variations characteristic for the first few normal modes.
Such selection was previously made \cite{Togashi2007} for scallop myosin
and $\mathrm{F}_{1}$-ATPase,
and it has been adopted in the present study for myosin V.
For KIF1A, the labels have been chosen in such a way that
motions in switch I and switch II regions are well resolved.

\subsection*{Profiles of elastic potential energy}

The collective mechanical coordinate $\phi$ along a relaxation path was
defined by requiring that its dynamics obeys the equation
$d\phi / dt = - \Gamma \partial U / \partial \phi$ and that
$\phi \to 0$ as $t \to \infty$.
Multiplying both parts of this equation by $d\phi / dt$,
we find that it is equivalent to the equation
\begin{equation}
\frac{d\phi}{dt} = - \sqrt{- \Gamma \frac{dU}{dt}}.
\label{coordinate}
\end{equation}
Equation (\ref{coordinate}) can be used to determine the
coordinate $\phi$ along a given relaxation trajectory and the dependence
of the elastic energy $U$ on this coordinate.

For each point along the trajectory, the time $t$ when it is reached in
the relaxation process is known.
Moreover, the actual network configuration corresponding to this point
is also known from the simulation.
Therefore, for each point specified by time $t$ the
respective elastic energy $U(t)$ is determined.
The mechanical coordinate $\phi$,
reached at time $t$, is given by the integral
\begin{equation}
\phi(t) = \int_{t}^{\infty} \sqrt{- \Gamma \frac{dU}{dt}} dt.
\end{equation}

\subsection*{Normal mode description}

We provide a summary of the results
on the normal mode description of conformational relaxation processes.
If deviations $\mathbf{r}_{i} = \mathbf{R}_{i} - \mathbf{R}_{i}^{(0)}$ from
the reference conformation are small for all particles,
the nonlinear equations (\ref{relaxation}),
describing conformational relaxation of an elastic network,
can be linearized:
\begin{equation}
\frac{d\mathbf{r}_{i}}{dt} = - \sum_{j=1}^{N} A_{ij} \left[ \mathbf{u}_{ij}^{(0)} \cdot \left( \mathbf{r}_{i}-\mathbf{r}_{j}\right) \right] \mathbf{u}_{ij}^{(0)},
\label{linear}
\end{equation}
where $\mathbf{u}_{ij}^{(0)} = \left( \mathbf{R}_{i}^{(0)}-\mathbf{R}_{j}^{(0)} \right) / d_{ij}^{(0)}$.

Equations (\ref{linear}) can be written in the matrix form as
\begin{equation}
\frac{d\mathbf{r}_{i}}{dt} = - \sum_{j=1}^{N} \mathbf{\Lambda}_{ij} \mathbf{r}_{j},
\label{linear2}
\end{equation}
where $\mathbf{\Lambda}$ is the $3N\times 3N$ linearization matrix:
\begin{eqnarray}
\Lambda_{i\sigma, j\eta} &=& A_{ij} u_{ij,\sigma}^{(0)} u_{ij,\eta}^{(0)} \ (\mathrm{for}\ i \neq j)\nonumber\\
\Lambda_{i\sigma, i\eta} &=& - \sum_{j} A_{ij} u_{ij,\sigma}^{(0)} u_{ij,\eta}^{(0)},
\label{linearmat}
\end{eqnarray}
where $\sigma,\ \eta = (x,\ y,\ z)$.

The general solution of these linear differential
equations is given by a superposition of $3N - 6$ exponentially decaying
normal modes, i.e.
\begin{equation}
\mathbf{r}_{i}(t) = \sum_{\alpha=1}^{3N-6} k_{\alpha} e^{- \Gamma \lambda_{\alpha} t} \mathbf{e}_{i}^{(\alpha)}.
\label{nmdecay}
\end{equation}
Here, $\lambda_{\alpha}$ and $\mathbf{e}_{i}^{(\alpha)}$ are the eigenvalues
and the eigenvectors of the linearization matrix, i.e.
\begin{equation}
\mathbf{\Lambda} \mathbf{e}_{i}^{(\alpha)} = \lambda_{\alpha} \mathbf{e}_{i}^{(\alpha)}.
\label{eigen}
\end{equation}
This matrix has $3N$ eigenvalues, but 6 of them must be zero,
corresponding to free translations and rotations of the entire network.

Generally, all normal modes are initially present.
As time goes on, first the normal modes with the larger eigenvalues
$\lambda_{\alpha}$ decay.
In the long time limit, relaxation is characterized
by the soft modes corresponding to the lowest eigenvalues.

Figure \ref{Fig-eigen} shows the computed eigenvalue spectra of myosin V and KIF1A.
The eigenvalues are normalized to the lowest nonzero eigenvalue
$\lambda_{1}$ and the logarithmic representation is chosen.

Note that in both motor proteins a significant gap,
separating the soft mode from the rest of the spectrum, is present.
This means that, in the linear approximation, long-time relaxation in
these proteins is effectively characterized by a single degree of
freedom, representing the amplitude of the soft mode. The pattern of
displacements of particles (i.e., residues) from the reference
positions is determined by the eigenvector $\mathbf{e}_{i}^{(1)}$ of
the soft mode.

In the plane $(u_{12},\ u_{13})$ of the distances between the
labels $1$, $2$ and $3$,
used by us for the visualization of conformational motions,
the exponential relaxation motion corresponding to the soft
mode should proceed along the direction defined by the vector with the
components
$\mathbf{u}_{12}^{(0)} \cdot (\mathbf{e}_{1}^{(1)} - \mathbf{e}_{2}^{(1)})$ and 
$\mathbf{u}_{13}^{(0)} \cdot (\mathbf{e}_{1}^{(1)} - \mathbf{e}_{3}^{(1)})$.
Such directions are indicated by dotted blue lines
in Figure \ref{Fig-Myo5}D and Figure \ref{Fig-KIF1A}D.

When relaxation is reduced to a single soft mode, the
elastic potential is quadratic in terms of the mechanical coordinate,
i.e. $U(\phi) = (1/2) \lambda_{1} \phi^{2}$.

Note that the representation of the relaxation process
as a superposition (\ref{nmdecay}) of normal relaxation modes
holds only in the harmonic domain, i.e. when linearization (\ref{linear})
of full nonlinear relaxation dynamics equations (\ref{relaxation}) is valid.
If dynamics is nonlinear and the linearization does not hold,
relaxation dynamics cannot be viewed {\em at all} as a superposition
of any normal modes.
Whether just one normal mode or many of them should be included
into a description of long-time relaxation dynamics is
determined by the properties of the eigenvalue spectrum and
not related to the possible invalidity
of the harmonic approximation.

As an extension, iterative normal mode analysis
has been proposed \cite{Miyashita,IterNMAFit}.
This method is applied to obtain an optimal sequence of
conformational states, transforming an initial given conformation
into a target conformation,
which may be known with a low resolution or only partially,
and thus to reconstruct missing details of that structure.
Each next conformation in the sequence is obtained by making a step
into the direction maximizing similarity with the target,
restricted however to a superposition of a certain number of
the lower normal modes.
At the next iteration step, the previous conformation is chosen
as a new reference state and a new set of normal modes is determined.
This prediction method is useful and provides valuable results,
e.g., in the refinement of low-resolution structures
from electron microscopy \cite{IterNMAFit}.
It should be however emphasized that the sequence of conformational
states yielded by such a method is generally different from the path
along which conformational relaxation from the target to the reference
state would proceed.
Even in the normal mode approximation, dynamics of conformational
relaxation depends not only on the eigenvectors, but also --- and
very significantly --- on the eigenvalues of normal modes.
Generally, the next iteration state in this method would not be
the next conformation along the actual relaxation path.
This difference can be clearly demonstrated by considering
the example of KIF1A.
The conformational relaxation path transforming the initial
ATP-bound state into the (equilibrium) ADP-bound state
is non-monotonous (Figure \ref{Fig-KIF1A}).
It proceeds via intermediate states (particularly of the
switch I region) which cannot be obtained by gradual interpolation
maximizing similarity of the structures along the optimization path. 

\section*{Acknowledgments}

We thank M. Ueda, H. Takagi and T. Komori for helpful comments.



\end{document}